\documentclass[aps,reprint,superscriptaddress,floatfix,showpacs,prl]{revtex4-2}
\usepackage[unicode=true,breaklinks=true,colorlinks=true]{hyperref}
\usepackage{graphicx}
\usepackage{amsmath}
\usepackage{amssymb}
\usepackage{amsfonts}
\usepackage{appendix}
\usepackage{newtxtext}
\usepackage{newtxmath}
\usepackage[all]{hypcap}
\usepackage{natbib}
\usepackage{multirow}
\usepackage{booktabs}
\usepackage{bbold}
\usepackage{bm}
\usepackage{bbm}
\usepackage[dvipsnames]{xcolor}
\usepackage{physics}
\usepackage{comment}
\usepackage[export]{adjustbox}
\usepackage{orcidlink}
\usepackage{balance}
\hypersetup{
    colorlinks=true,        
    urlcolor=darkgray,      
    citecolor=BrickRed,      
    linkcolor=RoyalPurple,     
    breaklinks=true         
}
\usepackage{pdfpages}
\makeatletter
\AtBeginDocument{\let\LS@rot\@undefined}
\makeatother
\begin{document}
\title{Observation of period doubling and higher multiplicities in a driven single-spin system}

%
%
\author{Dhruv Deshmukh\,\orcidlink{0009-0000-3779-3663}}
\thanks{Contact author: dhruv.deshmukh@uni-ulm.de}
\affiliation{Institute for Complex Quantum Systems, 
Universit\"{a}t Ulm, Albert-Einstein-Allee 11, D-89069 Ulm, Germany.}
\affiliation{Center for Integrated Quantum Science and Technology, 
Albert-Einstein-Allee 11, 89081 Ulm, Germany.}
\author{Ra\'ul B. Gonz\'alez\,\orcidlink{0000-0002-3507-120X}}
\thanks{Contact author: raul.gonzalez-brouwer@uni-ulm.de}
\affiliation{Institute for Quantum Optics, 
Universit\"{a}t Ulm, Albert-Einstein-Allee 11, 89081 Ulm, Germany.}
\affiliation{Center for Integrated Quantum Science and Technology, 
Albert-Einstein-Allee 11, 89081 Ulm, Germany.}
\author{Roberto Sailer\,\orcidlink{0009-0005-9039-9426}}
\thanks{Contact author: roberto.sailer@uni-ulm.de}
\affiliation{Institute for Quantum Optics, 
Universit\"{a}t Ulm, Albert-Einstein-Allee 11, 89081 Ulm, Germany.}
\affiliation{Center for Integrated Quantum Science and Technology, 
Albert-Einstein-Allee 11, 89081 Ulm, Germany.}
\author{Fedor Jelezko\,\orcidlink{0000-0001-5759-3917}}
\affiliation{Institute for Quantum Optics, 
Universit\"{a}t Ulm, Albert-Einstein-Allee 11, 89081 Ulm, Germany.}
\affiliation{Center for Integrated Quantum Science and Technology, 
Albert-Einstein-Allee 11, 89081 Ulm, Germany.}
\author{Ressa S. Said\,\orcidlink{0000-0003-4510-6397}}
\thanks{Present address: XeedQ GmbH, Wilhelm-Runge-Straße 10/3, 89081 Ulm, Germany.}
\affiliation{Institute for Quantum Optics, 
Universit\"{a}t Ulm, Albert-Einstein-Allee 11, 89081 Ulm, Germany.}
\affiliation{Center for Integrated Quantum Science and Technology, 
Albert-Einstein-Allee 11, 89081 Ulm, Germany.}
\author{Joachim Ankerhold\,\orcidlink{0000-0002-6510-659X}}
\thanks{Contact author: joachim.ankerhold@uni-ulm.de}
\affiliation{Institute for Complex Quantum Systems, 
Universit\"{a}t Ulm, Albert-Einstein-Allee 11, D-89069 Ulm, Germany.}
\affiliation{Center for Integrated Quantum Science and Technology, 
Albert-Einstein-Allee 11, 89081 Ulm, Germany.}


\begin{abstract}
One of the prime features of quantum systems strongly driven by external time-periodic fields is the subharmonic response with integer multiples of the drive period $k\, T_d$ due to long-lived interference. Here, we demonstrate experimentally, based on a careful theoretical analysis, period doubling and higher multiplicities ($k=2,\ldots 5$) for one of the most fundamental systems, namely, an individual spin $1/2$. Nitrogen-vacancy centers in diamond support sufficiently stable coherent dynamics owing  to long coherence times and allow for optical addressability of their spin states. This allows to monitor coherent period $k$-tupling oscillations over a broad set of driving parameters in the vicinity of the ideal manifolds. In this domain, superimposed low-frequency modulations serve as unique proxy for the approach toward period $k$-tupling.
\end{abstract}
\maketitle
\begin{figure*}[t]
    \centering
    \includegraphics[width=\linewidth]{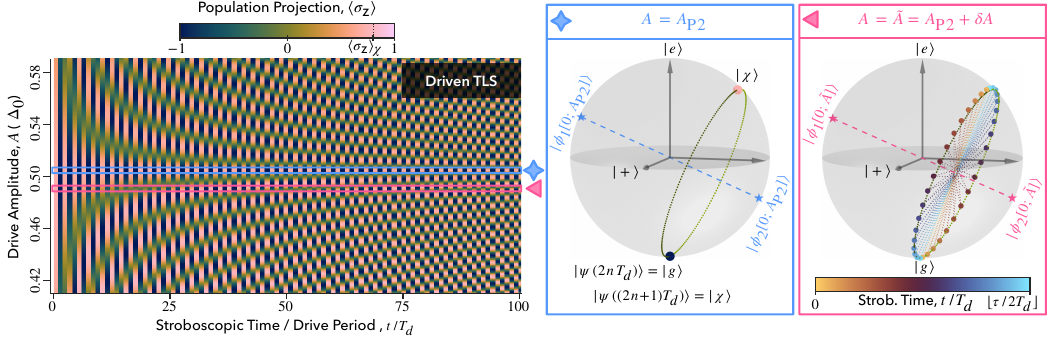}
    \caption{Stroboscopic population dynamics of a resonantly driven spin $1/2$ in multiples of the driving period $T_d$ according to Floquet theory. The system is  initialized in the ground state $|g\rangle$  and driving  amplitudes $A$ are close to the period doubling  amplitude $A_\mathrm{P2}$. Left panel: Population projection $\expval{\sigma_\mathsf{z}}$  evaluated after each drive cycle $T_d$. Right boxes: Dynamics on the Bloch sphere. At the amplitude $A_\mathrm{P2}$ (left, blue star), the alternating pattern of vertical blue-pink stripes shows that the projection flips between the initial value $-1$ and $\expval{\sigma_\mathsf{z}} {\chi}\simeq 0.7099$ and implies $2 T_d$-periodic revivals. On the Bloch sphere (right, blue box), this pattern corresponds to state $\ket{g}$ perpetually hopping to the state $\ket{\chi}$ and back. These states lie diametrically opposite on the circle about the associated Floquet mode axis $\ket{\phi_1[0;A_\mathrm{P2}]}-\ket{\phi_2[0;A_\mathrm{P2}]}$. Slightly away from the ideal amplitude, for $\tilde{A}=A_{P2}+\delta A$ (left, pink triangle), this alternating pattern is modulated. On the Bloch sphere (right, pink box), the stroboscopic trajectory lies on a circle passing through $\ket{g}$ on a plane perpendicular to a slightly shifted Floquet mode axis, $[\ket{\phi_1[0; \tilde{A}]}-\ket{\phi_2[0;\tilde{A}]}$. After each $2T_d$, the stroboscopic orbit shifts slightly,  returning to $|g\rangle$ only after a modulation period $\tau\propto 1/|\delta A|$.}
    \label{fig:TLS_PD}
\end{figure*}

Quantum systems can be controlled and manipulated when they are exposed to time-periodic external fields. Particularly fascinating is the interplay of non linearities and strong monochromatic driving that gives rise to a plethora of fundamental phenomena underpinned by the emergence of dressed states, Floquet states, or topological states far from equilibrium.

The arguably simplest non-linear quantum system is a single spin $1/2$ which has served as a paradigm for the physics of driven quantum systems as such. The conventional scenario involves a weak periodic drive for which the driving amplitudes are sufficiently smaller than the energy-level spacing. In this so-called Rabi regime, the linear dependence of the spin's frequency response on the drive amplitude provides a basis for pulse design and qubit gate operations. Much less common, but much richer, is the domain of strong periodic driving with drive amplitudes on the order of the transition frequency. Theory has formulated the general framework to describe this situation within Floquet theory \cite{kohler2005, Grifoni1998}, where quasienergies and Floquet states constitute the essential ingredients. Accordingly, the emergence of sidebands \cite{Phys.Rev.Lett._115_133601_2015}, localization \cite{grossmann1991,Miao2016}, Landau-Zener interferences \cite{Kayanuma1994,Sill2006}, and coherent-incoherent transitions \cite{magazzu2018} have been described and observed.

Here, we experimentally demonstrate and theoretically analyze one of the most intriguing features of driven non-linear quantum systems, namely, a subharmonic response due to frequency mixing, or equivalently, the breaking of discrete time-translational symmetry. For a two-level system (TLS), despite its simplicity and immense relevance for current quantum technologies, this feature has not been systematically explored yet, although previous work discussed indications \cite{Phys.Rev.B_92_054422_2015,Phys.Rev.Lett._115_133601_2015,Phys.Rev.A_95_053804_2017,Dykman2019}. We fill this gap by observing explicitly period doubling and higher multiplicities, thus revealing fundamental properties of the quantum mechanical time evolution. 

More specifically, a single TLS is shown to respond with coherent periodic dynamics with a period, $k\, T_d$, that is an integer multiple of the period, $T_d$, of the strong monochromatic field driving it. This appears generically, for any initial superposition of Floquet states, at any chosen initial time, and thus for any observable. We are able to monitor period doubling ($k=2$) and higher multiplicities ($k=3, 4, 5$)  in the vicinity of the respective ideal drive parameters by means of a stroboscopic measurement scheme. Our experimental findings verify a strongly enhanced response to mismatches and perturbations in drive amplitudes.

A pre-requisite for observing $k$-tupling with an individual spin is a sufficiently long coherence time since this response originates from revivals of interferences of the bare energy levels of the TLS. In this respect, experimentally, Nitrogen-Vacancy (NV) centers in diamond outperform other  TLS realizations thanks to their long coherence times at room temperature  \cite{jelezko2006single}, their field sensitivity \cite{taylor2008high,dolde2011electric}, and their optical addressability. For the periodic drive, no pulse engineering is required, and regular sinusoidal driving is sufficient. As prime indicator for the emergence of interferences corresponding to period doubling, we observe in the vicinity of ideal drive parameters a characteristic low-frequency modulation of the period-doubling response, whose period diverges upon approaching this manifold. Here, as shown experimentally, the TLS of the NV becomes particularly sensitive to any mismatches of amplitude/drive frequency from the $k$-tupling conditions.  An extended modeling captures the observed features very precisely. We also systematically retrieve higher multiplicities for various initial states. This way, we establish a consistent theoretical and experimental picture of a fundamental interference phenomenon for strongly driven systems. It specifically refers to the NV-center realization but conceptually applies to other TLS platforms as well. 

\emph{Theoretical background.\,\textbf{--}\,}
We start by considering the simplest situation of a TLS with level spacing $\Delta_0$ subject to a classical time-periodic drive with amplitude $A$, frequency $\omega_d=2\pi/T_d$, and initial phase $\varphi_0$, obeying the Hamiltonian,
\begin{equation}
\label{eq:hamiltonian}
    H = \frac{\Delta_0}{2} \sigma_\mathsf{z} + A \sin{(\omega_d t + \varphi_0)} \;\sigma_\mathsf{x}\,
\end{equation}
with Pauli matrices $\sigma_\mathsf{x}, \sigma_\mathsf{z}$.
The general theoretical framework to analyze the quantum dynamics of such a periodically driven system is based on the Floquet theory \cite{Grifoni1998, kohler2005, gramich2014, guo-phasespace}. It implies the existence of $T_d$-periodic solutions to Schr{\"o}dinger's equation, namely, the Floquet states $\mathrm{e}^{- \mathrm{i} \varepsilon_1 t / \hbar} |\phi_1(t)\rangle$ and $\mathrm{e}^{- \mathrm{i} \varepsilon_2 t / \hbar} |\phi_2(t)\rangle$, which serve as basis states for the quantum dynamics.   
Here, the quasienergies $\varepsilon_{1,2}$ are defined modulo the drive-photon energy $\hbar \omega_d$. Using these basis states, the stroboscopic dynamics of an arbitrary initial state $|\psi(0)\rangle$ can be written as
\begin{equation*}\label{eqn:StrobStateEvol}
    \ket{\psi(n T_d)} = \alpha\, \ket{\phi_1(0)} + \mathrm{e}^{- \mathrm{i} 2 \pi n \cdot (\vartriangle\varepsilon/\hbar \omega_d)} \beta\, \ket{\phi_2(0)} ,
\end{equation*}
where ${\vartriangle} \varepsilon = \varepsilon_2 - \varepsilon_1$ is the quasienergy difference. A particular  dynamical response, which we refer to as period $k$-tupling, occurs when ${\vartriangle} \varepsilon/\hbar \omega_d = j/k$ for some $j<k\in \mathbb{N}$. Despite the Hamiltonian and the Floquet states possessing $T_d$ periodicity, \textit{all} of the superposition states exhibit a lower symmetry with a nontrivial $k\, T_d$ periodicity \cite{SI}. In this situation, the quantum time evolution returns to the initial state $\ket{\psi(0)}$ only after $n=k$ driving periods $T_d$. On the dimensionless drive-parameter space $(\hbar\omega_d/\Delta_0-A/\Delta_0)$, the points for which the quasienergy differences satisfy the condition for period $k$-tupling form families of one-dimensional manifolds $\{\hbar\omega_{Pk}/\Delta_0, A_{Pk}/\Delta_0\}$. Below, we explore in detail the $k$-tupling dynamics of the driven TLS in the neighborhood of few isolated points on these manifolds, for $A \lesssim \Delta_0$ and $\hbar \omega_d = \Delta_0$. In this regime, the period $k$-tupling amplitudes for the Hamiltonian in Eq.\eqref{eq:hamiltonian} are given by $A_{Pk} \approx \Delta_0/k$.

Period $k$-tupling originates from revivals of quantum interference patterns at fractions of the drive frequency.  On the Bloch sphere, the transformation $\ket{\psi(t_0)} \to \ket{\psi(t_0 + n T_d)}$ in the case of period $k$-tupling can be seen as a sequence of discrete rotations about the Floquet mode axis by an angle $2 \pi j/k$. Specifically at the period-doubling amplitude $A_{P2}$, any state $\ket{\psi(t_0)}$ rotates by $\pi$ radians about the Floquet mode axis, $\ket{\phi_1(t_0)} - \ket{\phi_2(t_0)}$, after each period $T_d$ (see Fig.~\ref{fig:TLS_PD}). A small deviation in the drive amplitude $\tilde{A} = A_{P2} + \delta A$ results in a violation of the period-doubling condition, i.e., \  ${\vartriangle}\varepsilon/\hbar \omega_d = (1/2 + \xi), \abs{\xi}\ll 1$. The consequence of this deviation on the stroboscopic evolution is a \textit{slow rotation} along a circle around the Floquet axis as shown in Fig.~\ref{fig:TLS_PD}. It is this rotation that leaves a characteristic beating-like imprint on the experimental data and serves as a unique proxy for the approach toward the parameter manifold for period doubling. A numerical simulation of the stroboscopic population dynamics starting from the ground state of the bare TLS, equivalent to its survival probability, is depicted in Fig.~\ref{fig:TLS_PD}. It shows the population projection in the vicinity of the exact period-doubling amplitude $A_{P2} \simeq 0.5042 \Delta_0$ for $\hbar \omega_d = \Delta_0$. At the amplitude $A_{P2}$, one sees an alternating pattern of the projection values, whilst for small deviations ${A}\neq A_{P2}$, this ideal behavior is modulated with a period $\tau \propto 1/\delta A$ so that $\tau$ tends to infinity in the limit $A\to A_{P2}$, as illustrated in Fig.~\ref{fig:TLS_PD}. As we will see below, the actual experimental situation is  more intricate and requires an extended modeling but conceptually follows the described picture.

\emph{Experimental realization.\,\textbf{--}\,}
\begin{figure}[!t]
\centering
    \includegraphics[width=\columnwidth]{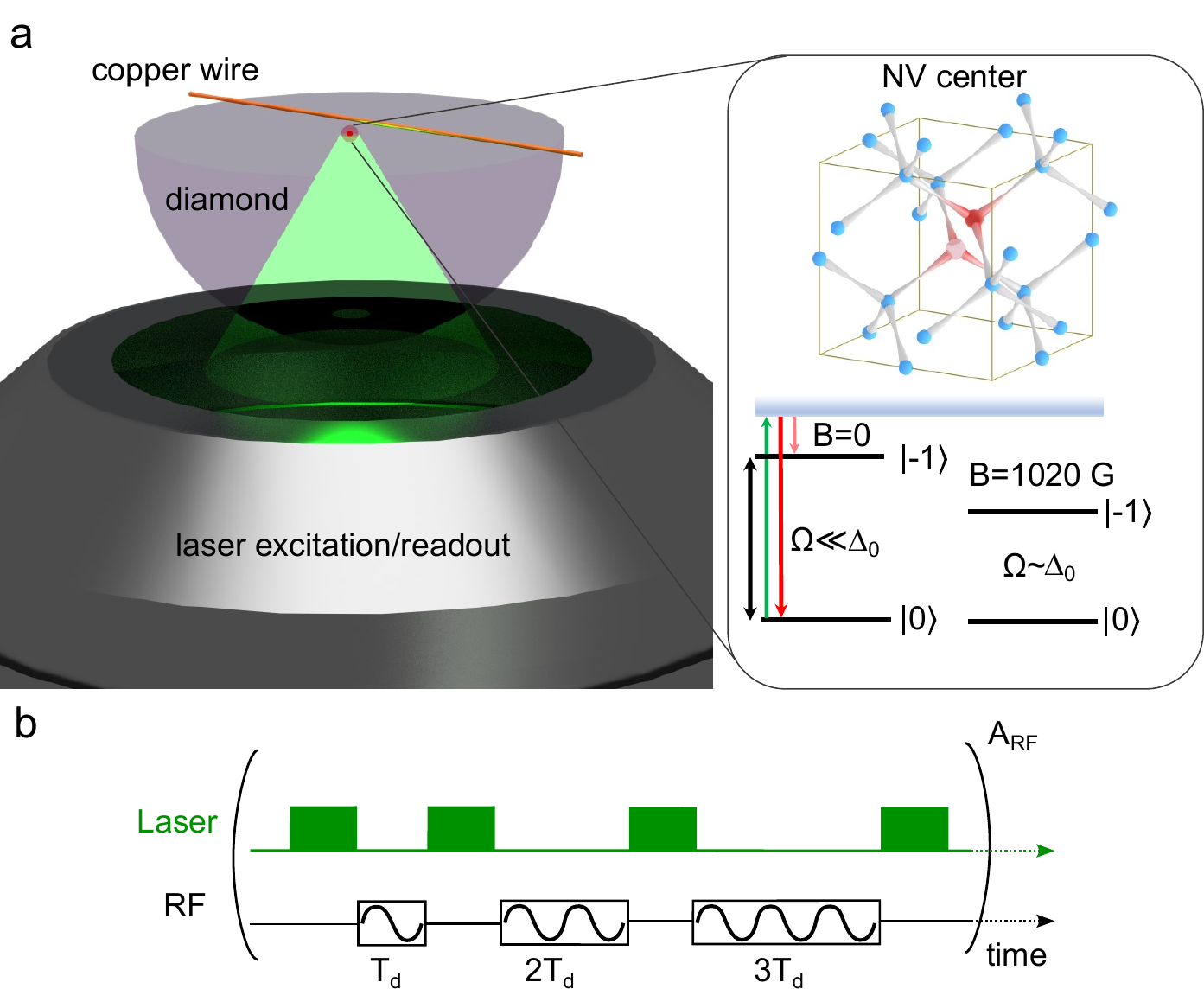}
    \caption{Experimental realization. (a) Setup, NV-center structure, and electronic energy levels. 
    A 532~nm laser (green cone; green arrow) is used for confocal microscopy experiments with a single NV center. The spin system is coherently controlled by an rf field via a copper wire antenna. The green laser is used to initialize the NV center into the~$\ket{m_S=0}$ spin state. 
    The spin state is read out by monitoring the fluorescence (red arrow) with different state-dependent intensities for $\ket{m_S=0}$ and $\ket{m_S=-1}$ (bright and dark red arrow). To reach~$A \sim \Delta_0$ for observing period $k$-tupling close to the ground-state level anticrossing, a high magnetic bias with an optimal alignment is applied.   
    (b) Driving and measurement protocol for observing period multiplicity. For each time trace, the sequence contains rectangular pulses with a fixed length~$T_\mathrm{pulse}=n T_d$~($T_d \equiv 2\pi/\omega_d\simeq 1/{\Delta_0}$). After each rf-pulse, a laser pulse is applied, where for this sequence a multiple shot of measurements is done. By using this fixed set of parameters, the rf-amplitude~$A$ is increased for each measurement cycle.}
    \label{Fig:energy}
\end{figure}
To demonstrate this dynamical response in a physical realization, we use a single NV center -- a point defect in the diamond lattice that contains a substitutional nitrogen atom and a neighboring vacancy~(Fig.~\ref{Fig:energy}). The ground state of the NV center's electronic energy levels~($^3$A) has a spin triplet~($S=1$). The state with zero spin projection,~$|m_S=0\rangle$ is separated from the degenerate states~$|m_S=\pm 1\rangle$ by a zero-field splitting of $\Delta_0=$~2.87~GHz. The effective TLS, spanned by~$\{|m_S=0\rangle, |m_S=-1\rangle\}$, can be achieved through an external static magnetic field splitting of $|m_S=\pm 1\rangle$ states. Our experiment uses a sample with solid immersion lens for a higher photon collection efficiency and a better signal-to-noise ratio. Spin initialization and read-out are performed by a~532~nm optical laser in~a home-built confocal setup. To control spin dynamics via radio-frequency fields, we utilize an arbitrary waveform generator with 0.02~ns~time resolution and 10-bit~voltage digitization which has~$\Delta V \approx$~0.5~mV at~$V_\mathrm{pp}=$~500~mV. It allows for scanning through the required amplitude ranges $A\sim \Delta_0$ for observing the period tupling. Our measured NV has an exceptional spin coherence time of $T^{\star}_2\approx$~75~$\mu$s which is over 100~times longer than the periodicity~$T_d=1/\Delta_0$. For the fulfillment of the constraint of a closed system, the longitudinal relaxation time $T_1 \approx 5 ~\mathrm{ms}$ by far exceeds $T^{\star}_2$. For more details on the experimental procedure, we refer to Ref.\cite{SI}.

Now, to make a strong driving protocol feasible, which is essential to observe period $k$-tupling, the spin transition must be tuned down sufficiently due to the power threshold of our microwave antenna, which lies in the range of megahertz. By adjusting the magnetic bias strength and orientation, the range $A\sim \Delta_0$ is close to the ground-state level anti-crossing (GSLAC) (Fig.~S19 in the Supplemental Material \cite{SI}). In order to achieve a high measurement contrast beneficial for high-fidelity state initialization~\cite{tetienne2012magnetic,jacques2009dynamic}, the field strength is set around 1020~G with an orientation aligned with the NV axis.  This yields a transition between the states $\ket{m_S=0}$ and $\ket{m_S=-1}$ of about $9.21$ MHz, on which the driving and measurement protocol is implemented as shown in Fig. \ref{Fig:energy}(b). The spin is initialized after 3~$\mu$s of optical pumping followed by a sequence of rectangular RF pulses with a pulse length~$T_\mathrm{pulse}=nT_d$ ($n\in\mathbb{N}$).  For a fixed $A_\mathrm{RF}$, the pulse length will be swept after each measurement, and the photoluminescence (PL) will be measured with the same initializing optical laser. The repetitive procedures for the various measurement times are controlled and processed by the suite~QUDI~\cite{binder2017qudi}.

\emph{Results and Discussion.\,\textbf{--}\,}
\begin{figure*}[!t]
    \centering
    \includegraphics[width=\textwidth]{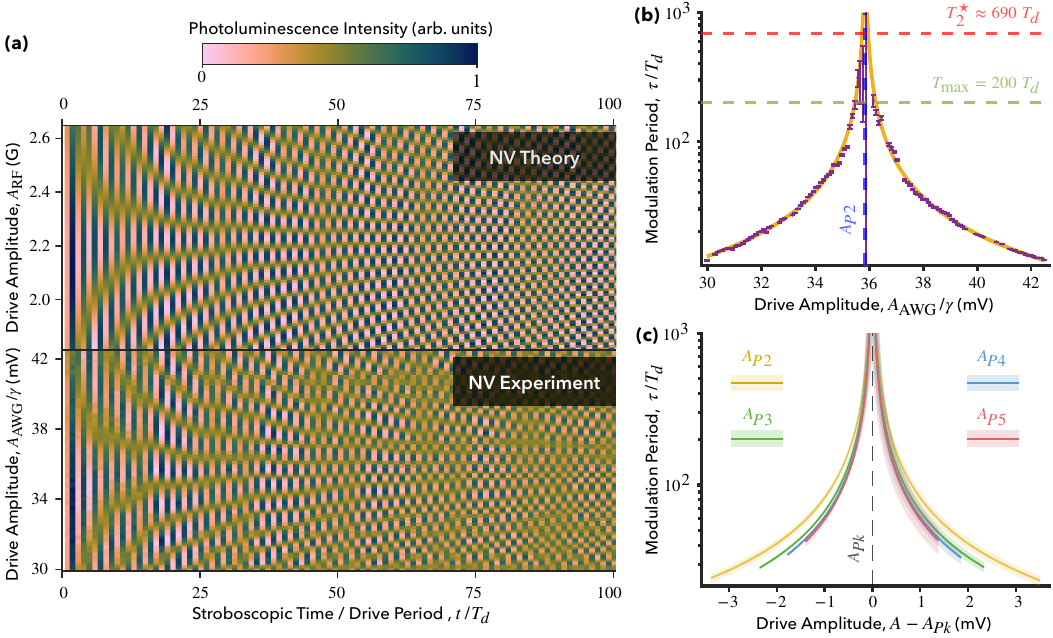}
    \caption{(a) Experimental data for photoluminescence and extended theoretical modeling for period doubling. The PL signal recorded after every drive cycle (bottom) is compared to that estimated through simulation (top). The theoretical simulations are obtained within an extended model \cite{SI} for driving the transition of $\Delta_0 = 8.00 $ MHz from $\ket{0,1/2}$ to $\ket{+}_\mu \equiv \mu \ket{-1,1/2} + (1-\abs{\mu}^2)^{1/2} \ket{0,-1/2}$ for $\abs{\mu}^2 = 0.9145$ ($D_\mathrm{ZFS} = 2.87$ GHz, $B_z = 1020.678$ G, and $\theta = 9^\circ$ for a weak misalignment of the magnetic field with respect to the NV axis). The mean level spacing for the experimental setup is $\expval{\Delta_0}$ = $9.21$ MHz. The drift in $\Delta_0$ is accounted for by a correction factor $\gamma$ (see Ref.\cite{SI}). (b) The modulation period for period doubling fits the hyperbolic curve predicted by the extended model; hyperbola $A>A_{P2}$: $(79.731 \pm 0.433)/|A-A_{P2}|$, hyperbola $A<A_{P2}$: $(75.447 \pm 0.424)/|A-A_{P2}|$ so that the amplitude for period doubling is extracted as $A_{P2}\mathrm{(exp)} = 35.844 \pm 0.018$ mV. (c) Same as panel (b) including higher multiplicities $k=2,3,4,5$ (for $\expval{\Delta_0}$ = $7.88$ MHz). See text and Ref.\cite{SI} for details.}
    \label{fig:NV_Results}
\end{figure*}
Figure \ref{fig:NV_Results}a (bottom) shows the stroboscopically measured PL signal after every drive cycle (see Ref.\cite{SI} for details).
One can clearly identify the characteristic period-doubling pattern consisting of alternate blue and pink vertical stripes along with its modulation manifested through the green fan-like structure, as already shown in Fig.~\ref{fig:TLS_PD}. 
As a salient feature, we extract in Fig.~\ref{fig:NV_Results}b the hyperbolic dependence of the modulation period $\tau \propto 1/|A-A_{P2}|$ on driving amplitude deviations, as predicted for the TLS. Not only does this enable us to locate the period-doubling amplitude precisely but it also corroborates the scaling of quasienergy differences with the drive amplitude in the vicinity of $A_{P2}$. The steep rise of the modulation period while approaching the amplitude $A_{P2}$ thus serves as a unique proxy for the emergence of period doubling. The corresponding strong susceptibility exposes minute details of the experimental setting that we discuss next.

A closer inspection of the predicted PL signal in Fig.\ref{fig:TLS_PD} and the measured one in Fig~\ref{fig:NV_Results}a (bottom) shows several discrepancies. First, and the most apparent, are the weak vertical fringes absent in the ideal TLS scenario. The latter one can only be explained through an extended treatment of the driven NV center that includes its close environment as shown in Fig.~~\ref{fig:NV_Results}a (top); when the NV spin is operated around the GSLAC, the neighboring $^{15}N$ nuclear spin not only causes a hyperfine splitting but also hybridizes with the electronic spin states. Furthermore, within the GSLAC region, the optical initialization procedure is known to promote polarization transfer, resulting in an efficient population of the state $|m_S=0, m_I=1/2\rangle$ \cite{Phys.Rev.Appl._6_064001_2016}. Thus, the TLS that is actually being driven with level spacing $\Delta_0$ turns out to consist of this state and the hybridized state $\ket{+}_\mu \equiv \mu\, \ket{-1,1/2} + \sqrt{1-|\mu|^2}\, \ket{0,-1/2}$ (cf. Ref.~\cite{SI}). Extracting the precise value of $\mu$ requires targeted experiments, however, the available data (level spacing between $\ket{0,1/2}$ and $\ket{+}_\mu$) allows us to estimate it to be $|\mu|^2 \sim 0.9$, via a comparison with a minimally extended model.  
Second, a careful analysis discloses a weak misalignment of the magnetic field with respect to the NV axis. Consequently, the Hamiltonian \eqref{eq:hamiltonian} needs to be further extended to include the longitudinal component of the driving; $\sigma_\mathsf{x}\to \cos(\theta)\sigma_\mathsf{x}+\sin(\theta)\sigma_\mathsf{z}$ with angle $\theta$ \cite{Phys.Rev.B_92_054422_2015}. Interestingly, it turns out that the angle under which the additional  fringes are tilted with respect to the vertical directly depends on $\theta$; this opens a convenient way to determine it. In the minimally extended model \cite{SI}, we find that $\theta\approx 9^\circ$  gives the best agreement between theory and experiment. 
Third, it appears that the level spacing $\Delta_0$ of the TLS slowly drifts between successive runs when sweeping the amplitude of the drive. To stay in tune ($\hbar\omega_d/\Delta_0=1$), the driving frequency $\omega_d$ is adapted accordingly, which is taken into account in the simulations.  
Fourth, the initialization fidelity of the electronic spin states after laser illumination is not perfect, typically in the range from 95$\%$ to 98$\%$ (see Refs. \cite{tetienne2012magnetic,robledo2011spin,pezzagna2021quantum}), which results in the relative contrast of fluorescence measurement. We estimate that this induces only very minor aberrations between theory and experiment. Finally, a weak decay of the PL contrast is ascribed to $T_2^*$. Based on this analysis, improved theoretical simulations \cite{SI} are depicted in Fig~\ref{fig:NV_Results} and show remarkable agreement with the experimental data. Future targeted experiments open up avenues to quantitatively extract the initial state population distribution, the branching ratio $|\mu|^2$, and thus precise values for hyperfine coupling constants as an alternative to the existing ODMR (optically detected magnetic resonance) and EPR (electron paramagnetic resonance) spectra. A further extended modeling including weak relaxation will then allow for even higher accuracy.

Is it possible to reveal in the response of the driven TLS also period $k$-tupling with $k>2$? In our experimental situation, this is indeed the case for $k$ = 3, 4, and 5 as explicitly demonstrated in Fig.~\ref{fig:NV_Results}c and discussed in Ref.\cite{SI}. Although the results were obtained with a coarser amplitude scan, the knowledge of the modulation period dependence on amplitude deviations from the ideal period $k$-tupling amplitude still allows us to identify $A_{Pk}$ with good precision. For all these cases, the hyperbola was found to fit $\tau\sim C_k/|A-A_{Pk}|$, wherein the coefficients $C_k$ decrease with increasing $k$ in agreement with theory. We emphasize that Figs.~\ref{fig:NV_Results}(b) and  ~\ref{fig:NV_Results}(c) collect data taken for different initial states which confirms the invariance of the $k$-tupling in agreement with theory.

\emph{Outlook.\,\textbf{--}\,} 
An individual spin $1/2$ realized in form of an NV center in a diamond crystal has been shown to exhibit period doubling and higher multiplicities when exposed to strong monochromatic fields. This phenomenon is a direct consequence of revivals of quantum interferences at fractions of the drive frequency and is observable due to the exceptional long coherence time of the NV sample. Within Floquet theory, we have provided a complete and quantitatively accurate understanding of this non-linear quantum dynamics in an extended model including relevant nuclear spins in the proximity of the NV spin. In the vicinity of the ideal drive parameters for $k$-tupling, the stroboscopic quantum dynamics shows low-frequency beatings. Their characteristic hyperbolic dependence on driving parameter mismatches serves as a proxy for the approach toward $k$-tupling.

From a fundamental perspective, these results provide the basis for a deeper understanding of the collective response of spin ensembles with growing size, related to the emergence of discrete time translational symmetry breaking \cite{Zhang2017,sacha2018,Smits2018,Choi2017}. Here, the case of multiple dipolar coupled NV centers surrounded by $^{13}C$ instead of a single NV center \cite{moon2024discrete} is an obvious scenario. On the applied side,  a natural next step is to analyze if, and if yes to what extent, the strongly driven NV spin turns into a very sensitive probe for other electromagnetic fields and/or residual spin degrees of freedom in its surrounding. Further, the non-trivial $k$-tupling dynamics suggests the possibility of engineering gate operations in quantum information processing with pulse times up to a few multiples of the bare transition frequency $\Delta_0\approx \hbar\omega_d$, i.e.,  much faster than those based on conventional Rabi pulses where the gate frequency is proportional to the drive amplitude $A\ll \Delta_0$. Conceptually, our findings directly apply to any other realization in atomic and solid-state physics with sufficient coherence times.\\
\nocite{Phys.Rev.B_79_075203_2009}

\noindent
\emph{Acknowledgments\,\textbf{--}\,}
D.D., R.S., R.S.S., F.J. and J.A. thank for financial support through the Baden-Württemberg Stiftung (QTBW-CDINQUA), R.S.S. and F.J. through the DFG, BMFTR (Co-GeQ, SPINNING, QRX), WM-BW (QC-4-BW), and ERC (Synergy Grant HyperQ), F.J., and J.A. through the IQST, BMFTR (QSENS-QMAT), and WM-BW.

D.D. developed the theoretical framework, proposed the experimental scheme, and analyzed the experimental data. R.B.G. and R.S. performed the experiments together. J.A. and R.S.S. coordinated the research collaboration. F.J. and J.A. secured funding and provided overall supervision. J.A., D.D., R.S.S., R.B.G., and R.S. contributed to manuscript writing. All authors participated in reviewing and editing the final version.
\newline\noindent
\emph{Data Availability\,\textbf{--}\,} The data that support the findings of this article are not publicly available. The data are available from the authors upon reasonable request.

%

\foreach \x in {1,...,27}
{%
\clearpage
\includepdf[pages={\x}]{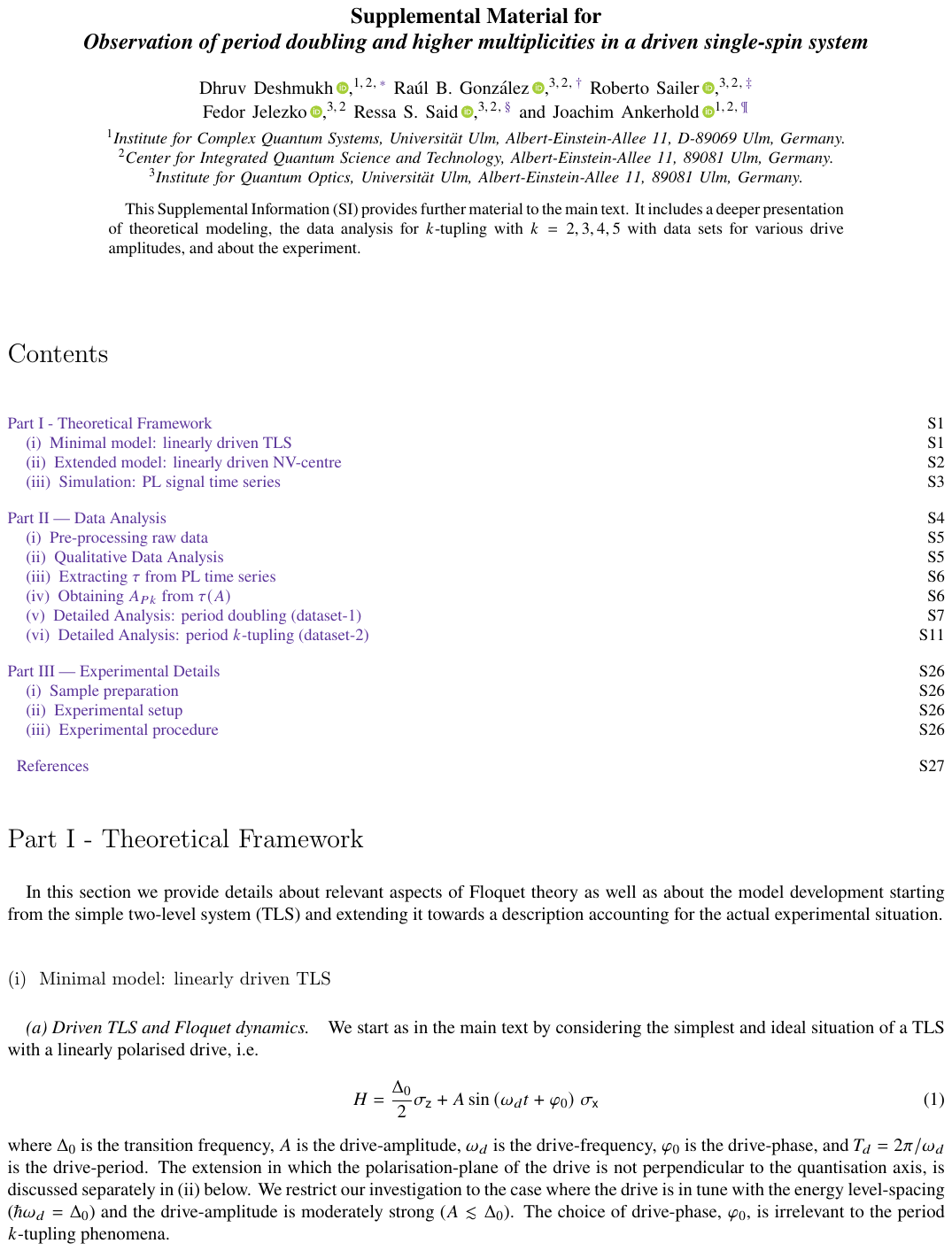} 
}

\begin{thebibliography}{29}%
\makeatletter
\providecommand \@ifxundefined [1]{%
 \@ifx{#1\undefined}
}%
\providecommand \@ifnum [1]{%
 \ifnum #1\expandafter \@firstoftwo
 \else \expandafter \@secondoftwo
 \fi
}%
\providecommand \@ifx [1]{%
 \ifx #1\expandafter \@firstoftwo
 \else \expandafter \@secondoftwo
 \fi
}%
\providecommand \natexlab [1]{#1}%
\providecommand \enquote  [1]{``#1''}%
\providecommand \bibnamefont  [1]{#1}%
\providecommand \bibfnamefont [1]{#1}%
\providecommand \citenamefont [1]{#1}%
\providecommand \href@noop [0]{\@secondoftwo}%
\providecommand \href [0]{\begingroup \@sanitize@url \@href}%
\providecommand \@href[1]{\@@startlink{#1}\@@href}%
\providecommand \@@href[1]{\endgroup#1\@@endlink}%
\providecommand \@sanitize@url [0]{\catcode `\\12\catcode `\$12\catcode `\&12\catcode `\#12\catcode `\^12\catcode `\_12\catcode `\%12\relax}%
\providecommand \@@startlink[1]{}%
\providecommand \@@endlink[0]{}%
\providecommand \url  [0]{\begingroup\@sanitize@url \@url }%
\providecommand \@url [1]{\endgroup\@href {#1}{\urlprefix }}%
\providecommand \urlprefix  [0]{URL }%
\providecommand \Eprint [0]{\href }%
\providecommand \doibase [0]{https://doi.org/}%
\providecommand \selectlanguage [0]{\@gobble}%
\providecommand \bibinfo  [0]{\@secondoftwo}%
\providecommand \bibfield  [0]{\@secondoftwo}%
\providecommand \translation [1]{[#1]}%
\providecommand \BibitemOpen [0]{}%
\providecommand \bibitemStop [0]{}%
\providecommand \bibitemNoStop [0]{.\EOS\space}%
\providecommand \EOS [0]{\spacefactor3000\relax}%
\providecommand \BibitemShut  [1]{\csname bibitem#1\endcsname}%
\let\auto@bib@innerbib\@empty
\bibitem [{\citenamefont {Kohler}\ \emph {et~al.}(2005)\citenamefont {Kohler}, \citenamefont {Lehmann},\ and\ \citenamefont {Hänggi}}]{kohler2005}%
  \BibitemOpen
  \bibfield  {author} {\bibinfo {author} {\bibfnamefont {S.}~\bibnamefont {Kohler}}, \bibinfo {author} {\bibfnamefont {J.}~\bibnamefont {Lehmann}},\ and\ \bibinfo {author} {\bibfnamefont {P.}~\bibnamefont {Hänggi}},\ }\bibfield  {title} {\bibinfo {title} {Driven quantum transport on the nanoscale},\ }\href@noop {} {\bibfield  {journal} {\bibinfo  {journal} {Phys. Rep.}\ }\textbf {\bibinfo {volume} {406}},\ \bibinfo {pages} {379} (\bibinfo {year} {2005})}\BibitemShut {NoStop}%
\bibitem [{\citenamefont {Grifoni}\ and\ \citenamefont {Hänggi}(1998)}]{Grifoni1998}%
  \BibitemOpen
  \bibfield  {author} {\bibinfo {author} {\bibfnamefont {M.}~\bibnamefont {Grifoni}}\ and\ \bibinfo {author} {\bibfnamefont {P.}~\bibnamefont {Hänggi}},\ }\bibfield  {title} {\bibinfo {title} {Driven quantum tunneling},\ }\href {https://doi.org/https://doi.org/10.1016/S0370-1573(98)00022-2} {\bibfield  {journal} {\bibinfo  {journal} {Physics Reports}\ }\textbf {\bibinfo {volume} {304}},\ \bibinfo {pages} {229} (\bibinfo {year} {1998})}\BibitemShut {NoStop}%
\bibitem [{\citenamefont {Deng}\ \emph {et~al.}(2015)\citenamefont {Deng}, \citenamefont {Orgiazzi}, \citenamefont {Shen}, \citenamefont {Ashhab},\ and\ \citenamefont {Lupascu}}]{Phys.Rev.Lett._115_133601_2015}%
  \BibitemOpen
  \bibfield  {author} {\bibinfo {author} {\bibfnamefont {C.}~\bibnamefont {Deng}}, \bibinfo {author} {\bibfnamefont {J.-L.}\ \bibnamefont {Orgiazzi}}, \bibinfo {author} {\bibfnamefont {F.}~\bibnamefont {Shen}}, \bibinfo {author} {\bibfnamefont {S.}~\bibnamefont {Ashhab}},\ and\ \bibinfo {author} {\bibfnamefont {A.}~\bibnamefont {Lupascu}},\ }\bibfield  {title} {\bibinfo {title} {Observation of floquet states in a strongly driven artificial atom},\ }\href {https://doi.org/10.1103/physrevlett.115.133601} {\bibfield  {journal} {\bibinfo  {journal} {Phys. Rev. Lett.}\ }\textbf {\bibinfo {volume} {115}},\ \bibinfo {pages} {133601} (\bibinfo {year} {2015})}\BibitemShut {NoStop}%
\bibitem [{\citenamefont {Grossmann}\ \emph {et~al.}(1991)\citenamefont {Grossmann}, \citenamefont {Dittrich}, \citenamefont {Jung},\ and\ \citenamefont {Hänggi}}]{grossmann1991}%
  \BibitemOpen
  \bibfield  {author} {\bibinfo {author} {\bibfnamefont {F.}~\bibnamefont {Grossmann}}, \bibinfo {author} {\bibfnamefont {T.}~\bibnamefont {Dittrich}}, \bibinfo {author} {\bibfnamefont {P.}~\bibnamefont {Jung}},\ and\ \bibinfo {author} {\bibfnamefont {P.}~\bibnamefont {Hänggi}},\ }\bibfield  {title} {\bibinfo {title} {Coherent destruction of tunneling},\ }\href@noop {} {\bibfield  {journal} {\bibinfo  {journal} {Phys. Rev. Lett.}\ }\textbf {\bibinfo {volume} {67}},\ \bibinfo {pages} {516} (\bibinfo {year} {1991})}\BibitemShut {NoStop}%
\bibitem [{\citenamefont {Miao}\ and\ \citenamefont {Zeng}(2016)}]{Miao2016}%
  \BibitemOpen
  \bibfield  {author} {\bibinfo {author} {\bibfnamefont {Q.}~\bibnamefont {Miao}}\ and\ \bibinfo {author} {\bibfnamefont {Y.}~\bibnamefont {Zeng}},\ }\bibfield  {title} {\bibinfo {title} {Coherent destruction of tunneling in two-level system driven across avoided crossing via photon statistics},\ }\href {https://doi.org/https://doi.org/10.1038/srep28959} {\bibfield  {journal} {\bibinfo  {journal} {Scientific Rep.}\ }\textbf {\bibinfo {volume} {6}},\ \bibinfo {pages} {28959} (\bibinfo {year} {2016})}\BibitemShut {NoStop}%
\bibitem [{\citenamefont {Kayanuma}(1994)}]{Kayanuma1994}%
  \BibitemOpen
  \bibfield  {author} {\bibinfo {author} {\bibfnamefont {Y.}~\bibnamefont {Kayanuma}},\ }\bibfield  {title} {\bibinfo {title} {Role of phase coherence in the transition dynamics of a periodically driven two-level system},\ }\href {https://doi.org/https://doi.org/10.1103/PhysRevA.50.843} {\bibfield  {journal} {\bibinfo  {journal} {Phys. Rev. A}\ }\textbf {\bibinfo {volume} {50}},\ \bibinfo {pages} {843} (\bibinfo {year} {1994})}\BibitemShut {NoStop}%
\bibitem [{\citenamefont {Sillanpää}\ \emph {et~al.}(2006)\citenamefont {Sillanpää}, \citenamefont {Lehtinen}, \citenamefont {Paila}, \citenamefont {Makhlin},\ and\ \citenamefont {Hakonen}}]{Sill2006}%
  \BibitemOpen
  \bibfield  {author} {\bibinfo {author} {\bibfnamefont {M.}~\bibnamefont {Sillanpää}}, \bibinfo {author} {\bibfnamefont {T.}~\bibnamefont {Lehtinen}}, \bibinfo {author} {\bibfnamefont {A.}~\bibnamefont {Paila}}, \bibinfo {author} {\bibfnamefont {Y.}~\bibnamefont {Makhlin}},\ and\ \bibinfo {author} {\bibfnamefont {P.}~\bibnamefont {Hakonen}},\ }\bibfield  {title} {\bibinfo {title} {Continuous- time monitoring of landau-zener interference in a cooper pair box},\ }\href {https://doi.org/https://doi.org/10.1103/PhysRevLett.96.187002} {\bibfield  {journal} {\bibinfo  {journal} {Phys. Rev. Lett.}\ }\textbf {\bibinfo {volume} {96}},\ \bibinfo {pages} {187002} (\bibinfo {year} {2006})}\BibitemShut {NoStop}%
\bibitem [{\citenamefont {Magazzù}\ \emph {et~al.}(2018)\citenamefont {Magazzù}, \citenamefont {Forn-Díaz}, \citenamefont {Belyansky}, \citenamefont {Orgiazzi}, \citenamefont {Yurtalan}, \citenamefont {Otto}, \citenamefont {A.}, \citenamefont {Wilson},\ and\ \citenamefont {Grifoni}}]{magazzu2018}%
  \BibitemOpen
  \bibfield  {author} {\bibinfo {author} {\bibfnamefont {L.}~\bibnamefont {Magazzù}}, \bibinfo {author} {\bibfnamefont {P.}~\bibnamefont {Forn-Díaz}}, \bibinfo {author} {\bibfnamefont {R.}~\bibnamefont {Belyansky}}, \bibinfo {author} {\bibfnamefont {J.-L.}\ \bibnamefont {Orgiazzi}}, \bibinfo {author} {\bibfnamefont {M.}~\bibnamefont {Yurtalan}}, \bibinfo {author} {\bibfnamefont {M.}~\bibnamefont {Otto}}, \bibinfo {author} {\bibfnamefont {L.}~\bibnamefont {A.}}, \bibinfo {author} {\bibfnamefont {C.}~\bibnamefont {Wilson}},\ and\ \bibinfo {author} {\bibfnamefont {M.}~\bibnamefont {Grifoni}},\ }\bibfield  {title} {\bibinfo {title} {Probing the strongly driven spin-boson model in a superconducting quantum circuit},\ }\href {https://doi.org/https://doi.org/10.1038/s41467-018-03626-w} {\bibfield  {journal} {\bibinfo  {journal} {Nature Comm.}\ }\textbf {\bibinfo {volume} {9}},\ \bibinfo {pages} {1403} (\bibinfo {year} {2018})}\BibitemShut {NoStop}%
\bibitem [{\citenamefont {Romh\'anyi}\ \emph {et~al.}(2015)\citenamefont {Romh\'anyi}, \citenamefont {Burkard},\ and\ \citenamefont {P\'alyi}}]{Phys.Rev.B_92_054422_2015}%
  \BibitemOpen
  \bibfield  {author} {\bibinfo {author} {\bibfnamefont {J.}~\bibnamefont {Romh\'anyi}}, \bibinfo {author} {\bibfnamefont {G.}~\bibnamefont {Burkard}},\ and\ \bibinfo {author} {\bibfnamefont {A.}~\bibnamefont {P\'alyi}},\ }\bibfield  {title} {\bibinfo {title} {Subharmonic transitions and bloch-siegert shift in electrically driven spin resonance},\ }\href {https://doi.org/10.1103/PhysRevB.92.054422} {\bibfield  {journal} {\bibinfo  {journal} {Phys. Rev. B}\ }\textbf {\bibinfo {volume} {92}},\ \bibinfo {pages} {054422} (\bibinfo {year} {2015})}\BibitemShut {NoStop}%
\bibitem [{\citenamefont {Rao}\ and\ \citenamefont {Suter}(2017)}]{Phys.Rev.A_95_053804_2017}%
  \BibitemOpen
  \bibfield  {author} {\bibinfo {author} {\bibfnamefont {K.~R.~K.}\ \bibnamefont {Rao}}\ and\ \bibinfo {author} {\bibfnamefont {D.}~\bibnamefont {Suter}},\ }\bibfield  {title} {\bibinfo {title} {Nonlinear dynamics of a two-level system of a single spin driven beyond the rotating-wave approximation},\ }\href {https://doi.org/10.1103/PhysRevA.95.053804} {\bibfield  {journal} {\bibinfo  {journal} {Phys. Rev. A}\ }\textbf {\bibinfo {volume} {95}},\ \bibinfo {pages} {053804} (\bibinfo {year} {2017})}\BibitemShut {NoStop}%
\bibitem [{\citenamefont {Dykman}(2019)}]{Dykman2019}%
  \BibitemOpen
  \bibfield  {author} {\bibinfo {author} {\bibfnamefont {M.~I.}\ \bibnamefont {Dykman}},\ }\bibfield  {title} {\bibinfo {title} {Coherent multiple-period states of periodically modulated qubits},\ }\href {https://doi.org/10.1103/PhysRevA.100.042101} {\bibfield  {journal} {\bibinfo  {journal} {Phys. Rev. A}\ }\textbf {\bibinfo {volume} {100}},\ \bibinfo {pages} {042101} (\bibinfo {year} {2019})}\BibitemShut {NoStop}%
\bibitem [{\citenamefont {Jelezko}\ and\ \citenamefont {Wrachtrup}(2006)}]{jelezko2006single}%
  \BibitemOpen
  \bibfield  {author} {\bibinfo {author} {\bibfnamefont {F.}~\bibnamefont {Jelezko}}\ and\ \bibinfo {author} {\bibfnamefont {J.}~\bibnamefont {Wrachtrup}},\ }\bibfield  {title} {\bibinfo {title} {Single defect centres in diamond: A review},\ }\href@noop {} {\bibfield  {journal} {\bibinfo  {journal} {physica status solidi (a)}\ }\textbf {\bibinfo {volume} {203}},\ \bibinfo {pages} {3207} (\bibinfo {year} {2006})}\BibitemShut {NoStop}%
\bibitem [{\citenamefont {Taylor}\ \emph {et~al.}(2008)\citenamefont {Taylor}, \citenamefont {Cappellaro}, \citenamefont {Childress}, \citenamefont {Jiang}, \citenamefont {Budker}, \citenamefont {Hemmer}, \citenamefont {Yacoby}, \citenamefont {Walsworth},\ and\ \citenamefont {Lukin}}]{taylor2008high}%
  \BibitemOpen
  \bibfield  {author} {\bibinfo {author} {\bibfnamefont {J.~M.}\ \bibnamefont {Taylor}}, \bibinfo {author} {\bibfnamefont {P.}~\bibnamefont {Cappellaro}}, \bibinfo {author} {\bibfnamefont {L.}~\bibnamefont {Childress}}, \bibinfo {author} {\bibfnamefont {L.}~\bibnamefont {Jiang}}, \bibinfo {author} {\bibfnamefont {D.}~\bibnamefont {Budker}}, \bibinfo {author} {\bibfnamefont {P.~R.}\ \bibnamefont {Hemmer}}, \bibinfo {author} {\bibfnamefont {A.}~\bibnamefont {Yacoby}}, \bibinfo {author} {\bibfnamefont {R.}~\bibnamefont {Walsworth}},\ and\ \bibinfo {author} {\bibfnamefont {M.~D.}\ \bibnamefont {Lukin}},\ }\bibfield  {title} {\bibinfo {title} {High-sensitivity diamond magnetometer with nanoscale resolution},\ }\href@noop {} {\bibfield  {journal} {\bibinfo  {journal} {Nature Physics}\ }\textbf {\bibinfo {volume} {4}},\ \bibinfo {pages} {810} (\bibinfo {year} {2008})}\BibitemShut {NoStop}%
\bibitem [{\citenamefont {Dolde}\ \emph {et~al.}(2011)\citenamefont {Dolde}, \citenamefont {Fedder}, \citenamefont {Doherty}, \citenamefont {N{\"o}bauer}, \citenamefont {Rempp}, \citenamefont {Balasubramanian}, \citenamefont {Wolf}, \citenamefont {Reinhard}, \citenamefont {Hollenberg}, \citenamefont {Jelezko} \emph {et~al.}}]{dolde2011electric}%
  \BibitemOpen
  \bibfield  {author} {\bibinfo {author} {\bibfnamefont {F.}~\bibnamefont {Dolde}}, \bibinfo {author} {\bibfnamefont {H.}~\bibnamefont {Fedder}}, \bibinfo {author} {\bibfnamefont {M.~W.}\ \bibnamefont {Doherty}}, \bibinfo {author} {\bibfnamefont {T.}~\bibnamefont {N{\"o}bauer}}, \bibinfo {author} {\bibfnamefont {F.}~\bibnamefont {Rempp}}, \bibinfo {author} {\bibfnamefont {G.}~\bibnamefont {Balasubramanian}}, \bibinfo {author} {\bibfnamefont {T.}~\bibnamefont {Wolf}}, \bibinfo {author} {\bibfnamefont {F.}~\bibnamefont {Reinhard}}, \bibinfo {author} {\bibfnamefont {L.~C.}\ \bibnamefont {Hollenberg}}, \bibinfo {author} {\bibfnamefont {F.}~\bibnamefont {Jelezko}}, \emph {et~al.},\ }\bibfield  {title} {\bibinfo {title} {Electric-field sensing using single diamond spins},\ }\href@noop {} {\bibfield  {journal} {\bibinfo  {journal} {Nature Physics}\ }\textbf {\bibinfo {volume} {7}},\ \bibinfo {pages} {459} (\bibinfo {year} {2011})}\BibitemShut {NoStop}%
\bibitem [{\citenamefont {Gramich}\ \emph {et~al.}(2014)\citenamefont {Gramich}, \citenamefont {Gasparinetti}, \citenamefont {Solinas},\ and\ \citenamefont {Ankerhold}}]{gramich2014}%
  \BibitemOpen
  \bibfield  {author} {\bibinfo {author} {\bibfnamefont {V.}~\bibnamefont {Gramich}}, \bibinfo {author} {\bibfnamefont {S.}~\bibnamefont {Gasparinetti}}, \bibinfo {author} {\bibfnamefont {P.}~\bibnamefont {Solinas}},\ and\ \bibinfo {author} {\bibfnamefont {J.}~\bibnamefont {Ankerhold}},\ }\bibfield  {title} {\bibinfo {title} {Lamb-shift enhancement and detection in strongly driven superconducting circuits},\ }\href@noop {} {\bibfield  {journal} {\bibinfo  {journal} {Phys. Rev. Lett.}\ }\textbf {\bibinfo {volume} {113}},\ \bibinfo {pages} {027001} (\bibinfo {year} {2014})}\BibitemShut {NoStop}%
\bibitem [{\citenamefont {Guo}(2021)}]{guo-phasespace}%
  \BibitemOpen
  \bibfield  {author} {\bibinfo {author} {\bibfnamefont {L.}\ \bibnamefont {Guo}},\ }\href {https://doi.org/10.1088/978-0-7503-3563-8} {\emph {\bibinfo {title} {Phase Space Crystals}}}\ (\bibinfo  {publisher} {Institute of Physics Publishing},\ \bibinfo {year} {2021})\BibitemShut {NoStop}%
\bibitem [{SI()}]{SI}%
  \BibitemOpen
  \href@noop {} {}\bibinfo {note} {See Supplemental Material below for the theoretical framework, extended NV-center modeling, numerical simulations, data-analysis and fitting procedures, additional period-$k$-tupling data for $k=2,3,4,5$, and experimental details, which includes Ref.~\cite{Phys.Rev.B_79_075203_2009}.}\BibitemShut {Stop}%
\bibitem [{\citenamefont {Tetienne}\ \emph {et~al.}(2012)\citenamefont {Tetienne}, \citenamefont {Rondin}, \citenamefont {Spinicelli}, \citenamefont {Chipaux}, \citenamefont {Debuisschert}, \citenamefont {Roch},\ and\ \citenamefont {Jacques}}]{tetienne2012magnetic}%
  \BibitemOpen
  \bibfield  {author} {\bibinfo {author} {\bibfnamefont {J.~P.}\ \bibnamefont {Tetienne}}, \bibinfo {author} {\bibfnamefont {L.}~\bibnamefont {Rondin}}, \bibinfo {author} {\bibfnamefont {P.}~\bibnamefont {Spinicelli}}, \bibinfo {author} {\bibfnamefont {M.}~\bibnamefont {Chipaux}}, \bibinfo {author} {\bibfnamefont {T.}~\bibnamefont {Debuisschert}}, \bibinfo {author} {\bibfnamefont {J.-F.}\ \bibnamefont {Roch}},\ and\ \bibinfo {author} {\bibfnamefont {V.}~\bibnamefont {Jacques}},\ }\bibfield  {title} {\bibinfo {title} {Magnetic-field-dependent photodynamics of single nv defects in diamond: an application to qualitative all-optical magnetic imaging},\ }\href@noop {} {\bibfield  {journal} {\bibinfo  {journal} {New Journal of Physics}\ }\textbf {\bibinfo {volume} {14}},\ \bibinfo {pages} {103033} (\bibinfo {year} {2012})}\BibitemShut {NoStop}%
\bibitem [{\citenamefont {Jacques}\ \emph {et~al.}(2009)\citenamefont {Jacques}, \citenamefont {Neumann}, \citenamefont {Beck}, \citenamefont {Markham}, \citenamefont {Twitchen}, \citenamefont {Meijer}, \citenamefont {Kaiser}, \citenamefont {Balasubramanian}, \citenamefont {Jelezko},\ and\ \citenamefont {Wrachtrup}}]{jacques2009dynamic}%
  \BibitemOpen
  \bibfield  {author} {\bibinfo {author} {\bibfnamefont {V.}~\bibnamefont {Jacques}}, \bibinfo {author} {\bibfnamefont {P.}~\bibnamefont {Neumann}}, \bibinfo {author} {\bibfnamefont {J.}~\bibnamefont {Beck}}, \bibinfo {author} {\bibfnamefont {M.}~\bibnamefont {Markham}}, \bibinfo {author} {\bibfnamefont {D.}~\bibnamefont {Twitchen}}, \bibinfo {author} {\bibfnamefont {J.}~\bibnamefont {Meijer}}, \bibinfo {author} {\bibfnamefont {F.}~\bibnamefont {Kaiser}}, \bibinfo {author} {\bibfnamefont {G.}~\bibnamefont {Balasubramanian}}, \bibinfo {author} {\bibfnamefont {F.}~\bibnamefont {Jelezko}},\ and\ \bibinfo {author} {\bibfnamefont {J.}~\bibnamefont {Wrachtrup}},\ }\bibfield  {title} {\bibinfo {title} {Dynamic polarization of single nuclear spins by optical pumping of nitrogen-vacancy color centers in diamond at room temperature},\ }\href {https://doi.org/10.1103/PhysRevLett.102.057403} {\bibfield  {journal} {\bibinfo  {journal} {Phys. Rev. Lett.}\ }\textbf {\bibinfo {volume} {102}},\ \bibinfo {pages} {057403}
  (\bibinfo {year} {2009})}\BibitemShut {NoStop}%
\bibitem [{\citenamefont {Binder}\ \emph {et~al.}(2017)\citenamefont {Binder}, \citenamefont {Stark}, \citenamefont {Tomek}, \citenamefont {Scheuer}, \citenamefont {Frank}, \citenamefont {Jahnke}, \citenamefont {M{\"u}ller}, \citenamefont {Schmitt}, \citenamefont {Metsch}, \citenamefont {Unden} \emph {et~al.}}]{binder2017qudi}%
  \BibitemOpen
  \bibfield  {author} {\bibinfo {author} {\bibfnamefont {J.~M.}\ \bibnamefont {Binder}}, \bibinfo {author} {\bibfnamefont {A.}~\bibnamefont {Stark}}, \bibinfo {author} {\bibfnamefont {N.}~\bibnamefont {Tomek}}, \bibinfo {author} {\bibfnamefont {J.}~\bibnamefont {Scheuer}}, \bibinfo {author} {\bibfnamefont {F.}~\bibnamefont {Frank}}, \bibinfo {author} {\bibfnamefont {K.~D.}\ \bibnamefont {Jahnke}}, \bibinfo {author} {\bibfnamefont {C.}~\bibnamefont {M{\"u}ller}}, \bibinfo {author} {\bibfnamefont {S.}~\bibnamefont {Schmitt}}, \bibinfo {author} {\bibfnamefont {M.~H.}\ \bibnamefont {Metsch}}, \bibinfo {author} {\bibfnamefont {T.}~\bibnamefont {Unden}}, \emph {et~al.},\ }\bibfield  {title} {\bibinfo {title} {Qudi: A modular python suite for experiment control and data processing},\ }\href@noop {} {\bibfield  {journal} {\bibinfo  {journal} {SoftwareX}\ }\textbf {\bibinfo {volume} {6}},\ \bibinfo {pages} {85} (\bibinfo {year} {2017})}\BibitemShut {NoStop}%
\bibitem [{\citenamefont {Broadway}\ \emph {et~al.}(2016)\citenamefont {Broadway}, \citenamefont {Wood}, \citenamefont {Hall}, \citenamefont {Stacey}, \citenamefont {Markham}, \citenamefont {Simpson}, \citenamefont {Tetienne},\ and\ \citenamefont {Hollenberg}}]{Phys.Rev.Appl._6_064001_2016}%
  \BibitemOpen
  \bibfield  {author} {\bibinfo {author} {\bibfnamefont {D.~A.}\ \bibnamefont {Broadway}}, \bibinfo {author} {\bibfnamefont {J.~D.~A.}\ \bibnamefont {Wood}}, \bibinfo {author} {\bibfnamefont {L.~T.}\ \bibnamefont {Hall}}, \bibinfo {author} {\bibfnamefont {A.}~\bibnamefont {Stacey}}, \bibinfo {author} {\bibfnamefont {M.}~\bibnamefont {Markham}}, \bibinfo {author} {\bibfnamefont {D.~A.}\ \bibnamefont {Simpson}}, \bibinfo {author} {\bibfnamefont {J.-P.}\ \bibnamefont {Tetienne}},\ and\ \bibinfo {author} {\bibfnamefont {L.~C.~L.}\ \bibnamefont {Hollenberg}},\ }\bibfield  {title} {\bibinfo {title} {Anticrossing spin dynamics of diamond nitrogen-vacancy centers and all-optical low-frequency magnetometry},\ }\href {https://doi.org/10.1103/PhysRevApplied.6.064001} {\bibfield  {journal} {\bibinfo  {journal} {Phys. Rev. Appl.}\ }\textbf {\bibinfo {volume} {6}},\ \bibinfo {pages} {064001} (\bibinfo {year} {2016})}\BibitemShut {NoStop}%
\bibitem [{\citenamefont {Robledo}\ \emph {et~al.}(2011)\citenamefont {Robledo}, \citenamefont {Bernien}, \citenamefont {Van Der~Sar},\ and\ \citenamefont {Hanson}}]{robledo2011spin}%
  \BibitemOpen
  \bibfield  {author} {\bibinfo {author} {\bibfnamefont {L.}~\bibnamefont {Robledo}}, \bibinfo {author} {\bibfnamefont {H.}~\bibnamefont {Bernien}}, \bibinfo {author} {\bibfnamefont {T.}~\bibnamefont {Van Der~Sar}},\ and\ \bibinfo {author} {\bibfnamefont {R.}~\bibnamefont {Hanson}},\ }\bibfield  {title} {\bibinfo {title} {Spin dynamics in the optical cycle of single nitrogen-vacancy centres in diamond},\ }\href@noop {} {\bibfield  {journal} {\bibinfo  {journal} {New Journal of Physics}\ }\textbf {\bibinfo {volume} {13}},\ \bibinfo {pages} {025013} (\bibinfo {year} {2011})}\BibitemShut {NoStop}%
\bibitem [{\citenamefont {Pezzagna}\ and\ \citenamefont {Meijer}(2021)}]{pezzagna2021quantum}%
  \BibitemOpen
  \bibfield  {author} {\bibinfo {author} {\bibfnamefont {S.}~\bibnamefont {Pezzagna}}\ and\ \bibinfo {author} {\bibfnamefont {J.}~\bibnamefont {Meijer}},\ }\bibfield  {title} {\bibinfo {title} {Quantum computer based on color centers in diamond},\ }\href@noop {} {\bibfield  {journal} {\bibinfo  {journal} {Applied Physics Reviews}\ }\textbf {\bibinfo {volume} {8}} (\bibinfo {year} {2021})}\BibitemShut {NoStop}%
\bibitem [{\citenamefont {Zhang}\ \emph {et~al.}(2017)\citenamefont {Zhang}, \citenamefont {Hess}, \citenamefont {Kyprianidis}, \citenamefont {Becker}, \citenamefont {Lee}, \citenamefont {Smith}, \citenamefont {Pagano}, \citenamefont {Potirniche}, \citenamefont {Potter}, \citenamefont {Vishwanath}, \citenamefont {Yao},\ and\ \citenamefont {Monroe}}]{Zhang2017}%
  \BibitemOpen
  \bibfield  {author} {\bibinfo {author} {\bibfnamefont {J.}~\bibnamefont {Zhang}}, \bibinfo {author} {\bibfnamefont {P.~W.}\ \bibnamefont {Hess}}, \bibinfo {author} {\bibfnamefont {A.}~\bibnamefont {Kyprianidis}}, \bibinfo {author} {\bibfnamefont {P.}~\bibnamefont {Becker}}, \bibinfo {author} {\bibfnamefont {A.}~\bibnamefont {Lee}}, \bibinfo {author} {\bibfnamefont {J.}~\bibnamefont {Smith}}, \bibinfo {author} {\bibfnamefont {G.}~\bibnamefont {Pagano}}, \bibinfo {author} {\bibfnamefont {I.-D.}\ \bibnamefont {Potirniche}}, \bibinfo {author} {\bibfnamefont {A.~C.}\ \bibnamefont {Potter}}, \bibinfo {author} {\bibfnamefont {A.}~\bibnamefont {Vishwanath}}, \bibinfo {author} {\bibfnamefont {N.~Y.}\ \bibnamefont {Yao}},\ and\ \bibinfo {author} {\bibfnamefont {C.}~\bibnamefont {Monroe}},\ }\bibfield  {title} {\bibinfo {title} {Observation of a discrete time crystal.},\ }\href {https://doi.org/10.1038/nature21413} {\bibfield  {journal} {\bibinfo  {journal} {Nature}\ }\textbf {\bibinfo {volume} {543}},\ \bibinfo
  {pages} {217} (\bibinfo {year} {2017})}\BibitemShut {NoStop}%
\bibitem [{\citenamefont {Sacha}\ and\ \citenamefont {Zakrzewsk}(2018)}]{sacha2018}%
  \BibitemOpen
  \bibfield  {author} {\bibinfo {author} {\bibfnamefont {K.}~\bibnamefont {Sacha}}\ and\ \bibinfo {author} {\bibfnamefont {J.}~\bibnamefont {Zakrzewsk}},\ }\bibfield  {title} {\bibinfo {title} {Time crystals: a review},\ }\href@noop {} {\bibfield  {journal} {\bibinfo  {journal} {Rep. Prog. Phys.}\ }\textbf {\bibinfo {volume} {81}},\ \bibinfo {pages} {016401} (\bibinfo {year} {2018})}\BibitemShut {NoStop}%
\bibitem [{\citenamefont {Smits}\ \emph {et~al.}(2018)\citenamefont {Smits}, \citenamefont {Liao}, \citenamefont {Stoof},\ and\ \citenamefont {van~der Straten}}]{Smits2018}%
  \BibitemOpen
  \bibfield  {author} {\bibinfo {author} {\bibfnamefont {J.}~\bibnamefont {Smits}}, \bibinfo {author} {\bibfnamefont {L.}~\bibnamefont {Liao}}, \bibinfo {author} {\bibfnamefont {H.~T.~C.}\ \bibnamefont {Stoof}},\ and\ \bibinfo {author} {\bibfnamefont {P.}~\bibnamefont {van~der Straten}},\ }\bibfield  {title} {\bibinfo {title} {Observation of a space-time crystal in a superfluid quantum gas},\ }\href {https://doi.org/10.1103/PhysRevLett.121.185301} {\bibfield  {journal} {\bibinfo  {journal} {Phys. Rev. Lett.}\ }\textbf {\bibinfo {volume} {121}},\ \bibinfo {pages} {185301} (\bibinfo {year} {2018})}\BibitemShut {NoStop}%
\bibitem [{\citenamefont {Choi}\ \emph {et~al.}(2017)\citenamefont {Choi}, \citenamefont {Choi}, \citenamefont {Landig}, \citenamefont {Kucsko}, \citenamefont {Zhou}, \citenamefont {Isoya}, \citenamefont {Jelezko}, \citenamefont {Onoda}, \citenamefont {Sumiya}, \citenamefont {Khemani}, \citenamefont {von Keyserlingk}, \citenamefont {Yao}, \citenamefont {Demler},\ and\ \citenamefont {Lukin}}]{Choi2017}%
  \BibitemOpen
  \bibfield  {author} {\bibinfo {author} {\bibfnamefont {S.}~\bibnamefont {Choi}}, \bibinfo {author} {\bibfnamefont {J.}~\bibnamefont {Choi}}, \bibinfo {author} {\bibfnamefont {R.}~\bibnamefont {Landig}}, \bibinfo {author} {\bibfnamefont {G.}~\bibnamefont {Kucsko}}, \bibinfo {author} {\bibfnamefont {H.}~\bibnamefont {Zhou}}, \bibinfo {author} {\bibfnamefont {J.}~\bibnamefont {Isoya}}, \bibinfo {author} {\bibfnamefont {F.}~\bibnamefont {Jelezko}}, \bibinfo {author} {\bibfnamefont {S.}~\bibnamefont {Onoda}}, \bibinfo {author} {\bibfnamefont {H.}~\bibnamefont {Sumiya}}, \bibinfo {author} {\bibfnamefont {V.}~\bibnamefont {Khemani}}, \bibinfo {author} {\bibfnamefont {C.}~\bibnamefont {von Keyserlingk}}, \bibinfo {author} {\bibfnamefont {N.~Y.}\ \bibnamefont {Yao}}, \bibinfo {author} {\bibfnamefont {E.}~\bibnamefont {Demler}},\ and\ \bibinfo {author} {\bibfnamefont {M.~D.}\ \bibnamefont {Lukin}},\ }\bibfield  {title} {\bibinfo {title} {Observation of discrete time-crystalline order in a disordered dipolar many-body
  system},\ }\href {https://doi.org/10.1038/nature21426} {\bibfield  {journal} {\bibinfo  {journal} {Nature}\ }\textbf {\bibinfo {volume} {543}},\ \bibinfo {pages} {221} (\bibinfo {year} {2017})}\BibitemShut {NoStop}%
\bibitem [{\citenamefont {Moon}\ \emph {et~al.}(2024)\citenamefont {Moon}, \citenamefont {Schindler}, \citenamefont {Smith}, \citenamefont {Druga}, \citenamefont {Zhang}, \citenamefont {Bukov},\ and\ \citenamefont {Ajoy}}]{moon2024discrete}%
  \BibitemOpen
  \bibfield  {author} {\bibinfo {author} {\bibfnamefont {L.~J.~I.}\ \bibnamefont {Moon}}, \bibinfo {author} {\bibfnamefont {P.~M.}\ \bibnamefont {Schindler}}, \bibinfo {author} {\bibfnamefont {R.~J.}\ \bibnamefont {Smith}}, \bibinfo {author} {\bibfnamefont {E.}~\bibnamefont {Druga}}, \bibinfo {author} {\bibfnamefont {Z.-R.}\ \bibnamefont {Zhang}}, \bibinfo {author} {\bibfnamefont {M.}~\bibnamefont {Bukov}},\ and\ \bibinfo {author} {\bibfnamefont {A.}~\bibnamefont {Ajoy}},\ }\bibfield  {title} {\bibinfo {title} {Discrete time crystal sensing},\ }\href@noop {} {\bibfield  {journal} {\bibinfo  {journal} {arXiv preprint arXiv:2410.05625}\ } (\bibinfo {year} {2024})}\BibitemShut {NoStop}%
\bibitem [{\citenamefont {Felton}\ \emph {et~al.}(2009)\citenamefont {Felton}, \citenamefont {Edmonds}, \citenamefont {Newton}, \citenamefont {Martineau}, \citenamefont {Fisher}, \citenamefont {Twitchen},\ and\ \citenamefont {Baker}}]{Phys.Rev.B_79_075203_2009}%
  \BibitemOpen
  \bibfield  {author} {\bibinfo {author} {\bibfnamefont {S.}~\bibnamefont {Felton}}, \bibinfo {author} {\bibfnamefont {A.~M.}\ \bibnamefont {Edmonds}}, \bibinfo {author} {\bibfnamefont {M.~E.}\ \bibnamefont {Newton}}, \bibinfo {author} {\bibfnamefont {P.~M.}\ \bibnamefont {Martineau}}, \bibinfo {author} {\bibfnamefont {D.}~\bibnamefont {Fisher}}, \bibinfo {author} {\bibfnamefont {D.~J.}\ \bibnamefont {Twitchen}},\ and\ \bibinfo {author} {\bibfnamefont {J.~M.}\ \bibnamefont {Baker}},\ }\bibfield  {title} {\bibinfo {title} {Hyperfine interaction in the ground state of the negatively charged nitrogen vacancy center in diamond},\ }\href {https://doi.org/10.1103/PhysRevB.79.075203} {\bibfield  {journal} {\bibinfo  {journal} {Phys. Rev. B}\ }\textbf {\bibinfo {volume} {79}},\ \bibinfo {pages} {075203} (\bibinfo {year} {2009})}\BibitemShut {NoStop}%
\end{thebibliography}
\end{document}